\documentstyle[12pt]{article}
\title{\Large\bf ON $q$-DEFORMED SPINNING \\ RELATIVISTIC PARTICLE}

\author{R.P.Malik
\thanks{E-mail:MALIK@THEOR.JINRC.DUBNA.SU}\\
      \small \it Bogoliubov Laboratory of Theoretical Physics,\\
 \small \it JINR, 141980 Dubna, Moscow Region, RUSSIA }

\begin{document}
\hoffset=-1truecm
\voffset=-2truecm
\baselineskip=16pt
\date{}
\maketitle
\begin{abstract}
A $q$-deformed free spinning relativistic particle is
discussed in the framework of the Lagrangian formalism.  Three equivalent
Lagrangians are obtained for this system which are endowed with
$q$-deformed local (super)gauge symmetries and reparametrization invariance.
It is demonstrated that these symmetries are on-shell equivalent only
for $ q = \pm1 $
under particular identification of the transformation parameters.
The same condition ($ q=\pm1 $) emerges due to the requirement that the
$q$-commutator of two supersymmetric gauge transformations should generate
a reparametrization plus a supersymmetric gauge
transformation. For a specific
gauge choice, the solutions for equations of motion respect
$GL_{\surd q}(1|1)$ and $GL_{q}(2)$
invariances for any arbitrary value of the evolution
parameter characterizing the quantum super world-line.
\end{abstract}
\pagenumbering{arabic}
\newpage
\noindent
There has been an upsurge of interest in the study of $q$-deformed
(so-called quantum) groups [1,2] during the past few years. These $q$-deformed
groups present examples of quasi-triangular Hopf algebras [3].
Together with the ideas of non-commutative geometry, it is expected that
the understanding of quantum groups might provide a {\it fundamental length}
[4] in the context of space-time quantization [5]. Thus, in addition to
Planck's
constant ($\hbar$) and the speed of light ($c$) that emerge from the study of
quantum mechanics and the special theory of relativity, the derivation of a
fundamental length in the context of quantum groups is conjectured to
complete the
{\it trio} of fundamental constants of nature.
Despite remarkable progress, the ideas of quantum groups have not percolated
to the level of multi-pronged physical applications. Few attempts have been
made, however, to provide some physical meaning to these
mathematical objects
in the context of concrete physical examples [6].
These quantum groups have
also been treated as gauge groups in an endeavour to develop $q$-deformed
gauge theories [7]. To obtain Lorentz covariant $q$-deformed gauge theories,
$q$-deformed path integral method, $q$-deformed field theories etc., it is
of utmost importance to develop Lagrangian formulations for some known
physical systems [8] in a cogent and consistent way.

The central theme of the present paper is to develop a Lagrangian formulation
for a $q$-deformed free spinning relativistic particle moving on a quantum
super world-line. We extend the prescription and methodology of Ref.[8]
to this system in a hope of developing a general scheme
for the discussion of other more complicated but realistic physical examples.
We obtain three equivalent $q$-deformed Lagrangians which are found to be
endowed with $q$-deformed local (super)gauge symmetries and reparametrization
invariance. These symmetries are shown to be equivalent on-shell {\it only}
for $ q=\pm 1$ under specific identification of
non-commuting gauge
parameters in terms of the commuting diffeomorphism parameter. The same
condition ($ q=\pm1 $) also emerges due to the requirement that the
$q$-commutator of two
supersymmetric gauge transformations must produce the sum of a
reparametrization and a supersymmetric gauge transformation as is essential
for pure supergravity theories.  One of the highlights of our work is the
existence of $GL_{\surd q}(1|1)$ and $GL_{q}(2)$ invariant solutions for
equations of motion under a specific gauge choice. This invariance persists at
any arbitrary value of the evolution parameter.  Since the emphasis in this
work is laid on symmetry considerations in the framework of the Lagrangian
formalism, we do not intend to discuss in detail the Hamiltonian formulation,
$\hbar$-quantization, $q$-deformed Dirac Brackets, etc., for this system which
is endowed with first-class as well as second-class constraints [9]. This
issue together with the $q$-deformed BRST formalism for a scalar as well as
a spinning particle will be reported elsewhere [10].

The simplest form of the local (super)gauge and reparametrization invariant
undeformed (classical) Lagrangian
that describes the
free motion ( $\dot p_{\mu} = 0$ ) of a massless relativistic spinning
particle is [11]
\begin{equation}
L_{F} = p_{\mu} \dot x^{\mu} + \frac{i}{2}\psi_{\mu}\;
\dot \psi^{\mu}
- \; \frac{e}{2} p^2 + i \chi \psi_{\mu} p^\mu, \label{1}
\end{equation}
where $x_{\mu}$, $p_{\mu}$, $e$ are {\it even} and $\psi_{\mu}$, $\chi$ are
{\it odd} elements of a Grassmann algebra. In the language of supergravity
theories, $x_{\mu}$ and $p_{\mu}$ are canonically conjugate target space
coordinates and momenta, $e$ and $\chi$ ($\chi^2 =0$) are Lagrange
multipliers that
are analogues of the vierbein and the Rarita-Schwinger fields and the
Lorentz vector $\psi_{\mu}$, which is the super partner of $x_{\mu}$,
presents
spin degrees of freedom and anticommutes with itself
( $\psi_{\mu} \psi_{\nu}+ \psi_{\nu}  \psi_{\mu} = 0$ ).
The "velocities"
$ \dot x_{\mu} \equiv \frac{d x^\mu}{d\tau} = ep^\mu - i \chi \psi^\mu $
and $\dot \psi^\mu \equiv \frac{d \psi^\mu}{d\tau}
= \chi p^\mu $ can be readily obtained from the above
Lagrangian where the trajectory of a spinning particle is
parametrized by an evolution variable $\tau$.
To obtain the $q$-analogue of the above Lagrangian we follow the discussion
of a $q$-deformed free relativistic
scalar particle [8] and generalize that prescription to
a $q$-deformed spinning
particle where the configuration space
corresponding to the Minkowski supermanifold
remains {\it flat} and {\it undeformed} ($ x_{\mu} x_{\nu}
= x_{\nu}  x_{\mu} ,\;
\psi_{\mu} \psi_{\nu}+ \psi_{\nu}  \psi_{\mu} = 0$) but the cotangent
supermanifold (momentum phase space) is $q$-{\it deformed} ($ x_{\mu} p_{\nu}
= q\; p_{\nu}  x_{\mu},
x_{\mu} x_{\nu} = x_{\nu}  x_{\mu},\; p_{\mu}p_{\nu}= p_{\nu} p_{\mu},
\psi_{\mu} \psi_{\nu}+ \psi_{\nu}  \psi_{\mu} = 0$)
in such a way that
the Lorentz invariance is respected for any arbitrary ordering of
$\mu$ and $\nu$. Here all the dynamical variables are taken
as hermitian elements of an algebra in involution ($ |q|= 1 $) and $q$ is a
non-zero $c$-number. As a consequence of the above deformation, the
following {\it on-shell} and {\it (graded)associative } $q$-(anti)commutation
relations emerge \footnote {These on-shell $q$-(anti)commutation relations
emerge from basic (un)deformed relations on a deformed cotangent supermanifold
and the equations of motion obtained from
the (un)deformed Lagrangians (1) {\it or} (11).
For instance, $\dot x_{\mu}\; \dot x_{\nu}= \dot x_{\nu}\; \dot x_{\mu}$
together with
$e\;p_{\mu}= q\; p_{\mu}\; e$ and $e\;\dot x_{\mu}= q\; \dot x_{\mu} e$
leads to the
relation
$ \chi( \psi_{\mu}\;p_{\nu} - \psi_{\nu}\;p_{\mu})
= q\;( p_\nu \chi\;\psi_{\mu} - p_\mu \chi\;\psi_{\nu})$. The requirement of
equality of this relation with the {\it similar one} emerging on-shell
from
$\dot \psi_{\mu}\; \psi_{\nu}+ \psi_{\nu}\; \dot \psi_{\mu}
+\dot \psi_{\nu}\; \psi_{\mu}+ \psi_{\mu}\; \dot \psi_{\nu}=0$ leads to:
$p_{\mu}\;\psi_{\nu}= q^{-1/2} \psi_{\nu}\;p_{\mu},
\chi\;p_{\mu}=q^{1/2} \;p_{\mu}\;\chi$ and
$\chi\;\psi_{\mu}=-q^{1/2} \;\psi_{\mu}\;\chi$ which are consistent with the
undeformed case in
the classical limit $ q \rightarrow 1 $.}
\begin{eqnarray}
&& x_{\mu}\;
x_{\nu}= x_{\nu}\; x_{\mu}, \quad \dot x_{\mu}\; \dot x_{\nu} = \dot x_{\nu}\;
\dot x_{\mu},\quad \dot x_{\mu} \;x_{\nu}= x_{\nu} \dot x_{\mu}, \quad
x_{\mu} \dot x_{\nu}= \dot x_{\nu} x_{\mu}, \nonumber\\ &&p_{\mu}\;p_{\nu}=
p_{\nu}\;p_{\mu},\quad  x_{\mu}\;p_{\nu} = q  p_{\nu}\; x_{\mu}, \quad \dot
x_{\mu}\; p_{\nu} = q\; p_{\nu}\;\dot x_{\mu}, \quad e\;x_{\mu} =
q\;x_{\mu}\;e, \nonumber\\ &&e\;p_{\mu} = q\;p_{\mu}\;e, \quad    e\;\dot
x_{\mu}\; = q\; \dot x_{\mu}\; e , \quad e\;\psi_{\mu} = q\;\psi_{\mu}\;e,
\quad e\;\chi= \chi\; e, \nonumber\\
&& \psi_{\mu}\; \psi_{\nu}+ \psi_{\mu} \psi_{\nu}=0,  \quad
\dot \psi_{\mu}\; \psi_{\nu}+ \psi_{\nu}\; \dot \psi_{\mu}=0,  \quad
\psi_{\mu}\; \dot \psi_{\nu}+ \dot\psi_{\nu}\; \psi_{\mu}=0, \nonumber\\
&&x_{\mu}\;\psi_{\nu}= q^{1/2} \psi_{\nu}\;x_{\mu},\quad
p_{\mu}\;\psi_{\nu}= q^{-1/2} \psi_{\nu}\;p_{\mu},\quad
p_{\mu}\;\dot \psi_{\nu}= q^{-1/2} \dot \psi_{\nu}\;p_{\mu}, \nonumber\\
&& \chi\;x_{\mu}=q^{1/2} \;x_{\mu}\;\chi,\qquad
\chi\;p_{\mu}=q^{1/2} \;p_{\mu}\;\chi,\qquad
\chi\;\psi_{\mu}=-q^{1/2} \;\psi_{\mu}\;\chi.
\label{2}
\end{eqnarray}
It is straightforward to see that in the limit when all
the {\it odd } Grassmann variables are set
equal to zero, we obtain $q$-commutation relations for a $q$-deformed
scalar free relativistic particle [8]. Furthermore, in the limit
$ q \rightarrow 1 $ the usual (anti)commutation relations among the
dynamical variables of the Lagrangian (1) emerge automatically. Consistent with
the $q$-(anti)commutation relations (2), the quantum super world-line,
traced out by the free motion of a spinning relativistic particle can be
defined in terms of the coordinate generator $x^\mu$ and the spin variable
$\psi^\mu$ as
\begin{eqnarray}
x_{\mu}(\tau)\; \psi^{\mu}(\tau)&=&
q^{1/2}\; \psi_{\mu}(\tau)\; x^{\mu}(\tau), \nonumber\\
(\psi^\mu(\tau))^2 &=& 0.
\label{3}
\end{eqnarray}
Here repeated indices are summed over (i.e. $ \mu= 0,1,2........D-1 $) and
the super world-line is parametrized by a real commuting variable $\tau$.
It is interesting to check that the quantum super world-line (3) remains
form-invariant under the following transformations
\begin{eqnarray}
x_{\mu} \; &\rightarrow&\; a\; x_{\mu} + \beta \psi_{\mu}, \nonumber\\
\psi_{\mu} \; &\rightarrow&\; \gamma x_{\mu} + d\; \psi_{\mu},
\label{4}
\end{eqnarray}
if we assume the (anti)commutativity of the variables
$\psi^\mu$ and $x^\mu$ with odd elements $\beta,\gamma$ and even elements
$ a,d $ of
a $2 \times 2 \quad  GL_{\surd q}(1|1)$ matrix obeying the braiding relations
in rows and columns as
\begin{eqnarray}
a\beta &=& q^{1/2} \beta a, \quad d\beta = q^{1/2} \beta d,
\quad \beta\;\gamma = - \gamma\;\beta, \quad \beta^2=\gamma^2=0, \nonumber\\
a\gamma &=& q^{1/2} \gamma a,\quad d\gamma = q^{1/2} \gamma d,\quad
ad-da=-(q^{1/2}-q^{-1/2})\beta \gamma.
\label{5}
\end{eqnarray}
It will be noticed that the $GL_{\surd q}(1|1)$ invariance is implied in
component pairs: $(x_{0},\psi_{0}).........(x_{D-1},\psi_{D-1})$,  namely;
\begin{equation}
\left(\begin{array}{c}
x_{i} \\
\psi_{i}\\ \end{array} \right) \quad
\rightarrow
\left(\begin{array}{cc}
a, & \beta \\
\gamma, & d \\ \end{array} \right) \quad
\left(\begin{array}{c}
x_{i} \\
\psi_{i}\\ \end{array} \right),
\label{6}
\end{equation}
for $i=0,1,2,3.......D-1$ separately . The other
candidate for the definition of the quantum super world-line:
$\psi_{\mu}p^\mu = q^{1/2} p_{\mu}\psi^\mu$ has not been taken
because $p_{\mu} \psi^\mu= 0$ is the constraint on the
system under consideration. Moreover, it can be readily seen that the
latter is contained in definition (3) due to the on-shell requirement
in $\dot x_{\mu} \psi^\mu= q^{1/2} \psi_{\mu} \dot x^\mu$ and equation (2).
It is
worth noting that the on-shell condition in
$ x_{\mu} \dot \psi^\mu= q^{1/2} \dot \psi_{\mu} x^\mu$ leads to the
definition of the $ GL_{q}(2)$ invariant quantum world-line
( $x_{\mu}(\tau)\; p^{\mu}(\tau) = q\; p_{\mu}(\tau)\; x^{\mu}(\tau)$)
taken in the
case of a free relativistic scalar particle [8] as it remains invariant
under
\begin{eqnarray}
x_{\mu} \; &\rightarrow&\; a\; x_{\mu} + b\; p_{\mu}, \nonumber\\
p_{\mu} \; &\rightarrow&\; c\; x_{\mu} + d\; p_{\mu},
\label{7}
\end{eqnarray}
if we assume the commutativity of the phase variables $ x^\mu$ and $p^\mu$
with the
elements $a,\;b,\;c$, and $d$  of a $ 2 \times 2 \quad GL_{q}(2)$
matrix obeying
following relationship
\begin{eqnarray}
ab &=& qba, \quad ac = qca, \quad cd = q dc,\quad bd=q db, \nonumber\\
bc &=& cb,\qquad ad - da =\;(q-q^{-1}) \; bc.
\label{8}
\end{eqnarray}
In the definition of the quantum world-line for a scalar particle, once
again, repeated indices are summed over and the $GL_{q}(2)$ invariance is
implied in the component pairs of the phase variables.
It is gratifying to see that the $GL_{q}(2)$ invariant quantum world-line
for a scalar relativistic particle emerges on-shell from the
$GL_{\surd q}(1|1)$ invariant quantum super world-line (3) for a
spinning relativistic particle.

The first-order Lagrangian, describing the free motion ($\dot p_{\mu}=0$)
of a massless $q$-deformed relativistic spinning particle, is
\begin{equation}
L_{f} = q^{1/2} p_{\mu} \dot x^{\mu} + \frac{i}{2}\psi_{\mu}\;
\dot \psi^{\mu}
- \; \frac{e}{1+ q^2} p^2 + i \chi \psi_{\mu} p^\mu, \label{9}
\end{equation}
where the $q^{1/2}$ factor appears in the first term due to the Legendre
transformation with $q$-symplectic metrices [8]
\begin{equation}
\Omega_{AB}(q)= \left(\begin{array}{cc}
0, & -q^{-1/2} \\
q^{1/2}, & 0 \\ \end{array} \right) \qquad  \mbox{and} \qquad
\Omega^{AB}(q) = \left(\begin{array}{cc}
0, & q^{-1/2} \\
-q^{1/2}, & 0 \\ \end{array} \right).
\label{10}
\end{equation}
In the last term of the Lagrangian (9), the variables $p^\mu$ and
$ \chi \psi^\mu$ are
arranged in such a way that for the differentiation with respect to $p^\mu$,
one
can exploit the $GL_{q}(2)$ invariant
( $(\chi \psi_{\mu})(\tau)\; p^{\mu}(\tau)
= q\; p_{\mu}(\tau)\; (\chi \psi^{\mu})(\tau)$)
differential calculus [12].
Here the $q$-Hamiltonian for a free spinning particle
has been taken to be:
$ H = \frac{e}{1+ q^2} p^2 - i \chi \psi_{\mu} p^\mu$.
One can include the mass term in the Lagrangian
(9) by invoking another Lorentz scalar $q$-anticommuting spinor variable
$\psi_{5}$ ( $ (\psi_{5})^2 = -1 $) as
\begin{equation}
L_{f}^m = q^{1/2} p_{\mu} \dot x^{\mu} + \frac{i}{2}(\psi_{\mu}\;
\dot \psi^{\mu} - \psi_{5} \dot \psi_{5})
- \; \frac{e}{1+ q^2} (p^2 - m^2)
+ i \chi (\psi_{\mu} p^\mu - \psi_{5} m ), \label{11}
\end{equation}
where the $\tau$ independent mass term ($m$) obeys the following
$q$-commutation relations with the rest of the dynamical variables
\begin{eqnarray}
e\;m &=& q m\;e, \quad \dot x_{\mu}\; m = q\; m\;\dot x_{\mu},
\quad x_{\mu}m\;=  q\; m\;x_{\mu}, \quad p_{\mu}\; m = m\; p_{\mu},
\nonumber\\
\chi m &=& q^{1/2} m \chi, \quad \psi_{\mu} m = q^{1/2} m \psi_{\mu},
\quad \psi_{5} m = q^{1/2} m \psi_{5}.
\label{12}
\end{eqnarray}
The $q$-(anti)commutation relations of $\psi_{5}$ with the rest of the
dynamical variables are the {\it same} as that of $\psi_{\mu}$ and both of
these anticommute ($ \psi_{\mu} \psi_{5} + \psi_{\mu} \psi_{5} = 0$).
The equations of motion from the Lagrangian (11) are
\begin{eqnarray}
\dot x_{\mu} &=& q^{1/2} ( e p_{\mu} - i \chi \psi_{\mu} )\nonumber\\
\dot \psi_{\mu}&=& q^{1/2} \chi p_{\mu}, \quad \dot p_{\mu} = 0, \nonumber\\
\dot \psi_{5}&=& q^{1/2} \chi m,
\label{13}
\end{eqnarray}
which satisfy the on-shell $q$-(anti)commutation relations (2) and (12).
In the differentiation of the Lagrangian (11) with respect to $\dot x_{\mu}$
and $p_{\mu}$, we have exploited the $GL_{q}(2)$ invariant differential
calculus developed in Ref.[12]. For instance, for $xy = q yx$ ,
any  monomial is arranged in the form $y^r\; x^s$ and then we use
\begin{eqnarray}
\frac{\partial ( y^r \;  x^s )}{\partial  x}
&=&\; y^{r}\; x^{s-1}\;q^r \; \frac{( 1 - q^{2s})}{( 1 - q^{2})},
\nonumber\\
\frac{\partial ( y^r \;  x^s )}{\partial  y}
&=&\; y^{r-1}\; x^{s}\; \frac{( 1 - q^{2r})}{( 1 - q^{2})},
\label{14}
\end{eqnarray}
where  $ r,s \in {\cal Z} $ are whole numbers (not fractions). For
differentiations with respect to the odd Grassmann variables $ \psi_{\mu},
\psi_{5}, \dot \psi_{\mu}, \dot \psi_{5}, \chi $, these variables are first
brought to the left side by using
$q$-(anti)commutation relations (2) and (12) in the corresponding
expressions, and then, differentiation
is carried out. Using the contravariant metric
of equation (10) and the Hamiltonian ($H$), one can check that equations
(13) can be rewritten as $ \dot x_{\mu}= \Omega^{AB}\partial_{A} x_{\mu}
\partial_{B} H \equiv q^{1/2} ( e p_{\mu} - i \chi \psi_{\mu} ),\;
\dot p_{\mu} = 0 \;\mbox{and} \;\dot \psi_{\mu} = -i \frac{\partial H}
{\partial \psi^{\mu}}$.

The second-order Lagrangian ($L_{s}^m$), describing the motion of a spinning
particle on the tangent manifold (velocity phase space), can be obtained
from the first-order
Lagrangian (11) by exploiting equations (2), (12) and (13) as given below:
\begin{equation}
L_{s}^m = \; \frac{q^2}{1 + q^2}\; e^{-1} \; ( \dot x_{\mu}
+ q^{1/2} i \chi \psi_{\mu})^2
+ \; \frac{e}{1 + q^2}\;  m^2  + \frac{i}{2}(\psi_{\mu}\;
\dot \psi^{\mu} - \psi_{5} \dot \psi_{5})
- i \chi \psi_{5} m.
\label{15}
\end{equation}
The consistent expression for the
canonical
momenta ($ p_{\mu} $) for the first- and second-order Lagrangians (11) and
(15)
\begin{equation}
p_{\mu}\; = \; q^{-3/2} \Bigl ( \frac{\partial L_{(f,s)}^m}{\partial
\dot x^{\mu}} \Bigr ) \equiv q^{-1/2}\;e^{-1}\;
(\dot x_{\mu} +q^{1/2} i \chi \psi_{\mu}),
\label{16}
\end{equation}
leads to its square as:
\begin{equation}
p_{\mu}\; p^{\mu} = e^{-2} ( \dot x_{\mu} +q^{1/2} i \chi \psi_{\mu})^2,
\label{17}
\end{equation}
if we use the $q$-(anti)commutation relations (2). To see that the
right hand side of equation (17) is the square of the mass, one has to
exploit the $GL_{q}(2)$ invariant differential calculus to differentiate
Lagrangian $L_{s}^m$ with respect to $e$. For instance, the first term of (15)
has to be first recast as $ \frac{( \dot x_{\mu} +q^{1/2} i \chi
\psi_{\mu})^2 e^{-1}}{ 1+ q^2}$ and then, differentiation with respect to $e$
has to be performed. The final outcome
\begin{equation}
\frac{q^4}{1 + q^2}
\Bigl [ m^2 \; - e^{-2}\;( \dot x_{\mu} +q^{1/2} i \chi \psi_{\mu})^2 \Bigr ]
= 0, \label{18}
\end{equation}
leads to the mass-shell condition for the
$q$-deformed spinning relativistic particle  as:
\begin{equation}
p_{\mu}\;p^{\mu} - m^2 = 0. \label{19}
\end{equation}

This equation is one of the
Casimir invariants of the Poincar\'{e} group corresponding to the undeformed
flat Minkowski space-time and it turns up here as the constraint condition.
The other constraint condition $ p_{\mu} \psi^\mu - m \psi_{5}= 0  $ appears
because of the differentiation with respect to $\chi$ in the first-order
Lagrangian (11) (which can be checked to be true for the Lagrangian (15) as
well).  These constraint conditions are in agreement with the discussion of
the Klein-Gordon equation and the Dirac-equation derived from the
representation theory of the Lorentz group and $q$-deformation of the
Dirac-$\gamma$ matrices [13].  To obtain the Lagrangian with a square root, it
is essential to succinctly express ($ e, e^{-1}$) in terms of the mass ($m$)
and
the square root of $ ( \dot x_{\mu} +q^{1/2} i \chi \psi_{\mu})^2$. It is not
straightforward to obtain ($e,e^{-1}$) from $m^2 = e^{-2} ( \dot x_{\mu}
+q^{1/2} i \chi \psi_{\mu})^2$ because of the non-commutativity of velocity,
mass, $\chi$ and $\psi_{\mu}$.  A nice and simple way to compute these is to
first start with
\begin{equation}
e^{-1} = \; f(q) \; m \;
[(\dot x_{\mu} +q^{1/2} i \chi \psi_{\mu})^2 ]^{-1/2},  \label{20}
\end{equation}
and insert it into (15) using $q$-(anti)commutation relations (2) and
(12) such that $e^{-1}$ and $e$ occupy various positions
in its first and second
terms. The requirement of
the equality of the resulting Lagrangians leads to
\footnote { We have used equation (20) and
$[( \dot x_{\mu} +q^{1/2} i \chi \psi_{\mu})^2]^{1/2} \; m
= f^2(q) \; m \;[( \dot x_{\mu} +q^{1/2} i \chi \psi_{\mu})^2]^{1/2} $ in all
three possible expressions in which both the terms can be recast.
The term-by-term equality for all the possibilities yields a common condition
$ f^4(q) = q^2 $.}
\begin{equation}
 f^4(q) = q^2. \label{21}
\end{equation}
This requirement is satisfied by  four values of
$f(q)$, namely; $ f(q) = \pm q^{1/2}; \pm i q^{1/2}$. The
key requirement, however, that the $q$-deformed Lagrangian should reduce to
the usual undeformed (classical) Lagrangian in the limit $ q \rightarrow 1$
restricts $f(q)$ to picking up {\it only} the value $ q^{1/2}$. Ultimately the
Lagrangian ($ L_{0}^m$)
with the square root turns out to be
\begin{equation}
L_{0}^m = q^{1/2}  m [(\dot x_{\mu} +q^{1/2} i \chi \psi_{\mu})^2 ]^{1/2}
+ \frac{i}{2}(\psi_{\mu}\;\dot \psi^{\mu} - \psi_{5} \dot \psi_{5})
- i \chi \psi_{5} m.
\label{22}
\end{equation}
All the three Lagrangians (11), (15) and (22) are equivalent
\footnote {The last two-terms in (22)
can be combined together to yield a more concise expression
$\frac{i}{2}(\psi_{\mu}\;\dot \psi^{\mu} + \psi_{5} \dot \psi_{5})$ using
the equation of motion $\dot \psi_{5}= q^{1/2} \chi \; m $. However, there are
certain subtleties in the proof of equivalence of the resulting Lagrangian
with the other two [9].}
as far
as symmetry properties are concerned. They differ drastically, however,
in the limit $ m \rightarrow 0 $.

The expression for the canonical momenta (16) is true for the
Lagrangian (22) with square root as well. To see this, we require
the $GL_{q}(2)$ invariant
relation
\begin{equation}
[(\dot x_{\mu} +q^{1/2} i \chi \psi_{\mu})^2 ]^{1/2} \; m
= q\;  m [(\dot x_{\mu} +q^{1/2} i \chi \psi_{\mu})^2 ]^{1/2},
\label{23}
\end{equation}
that emerges from (20) and the equality $ e^{-2}
= m^2  [(\dot x_{\mu} +q^{1/2} i \chi \psi_{\mu})^2 ]^{-1}$.
Finally we obtain,
\begin{equation}
p_{\mu}\; = \; q^{-3/2} \Bigl ( \frac{\partial L_{0}^m}{\partial
\dot x^{\mu}} \Bigr ) \equiv (\dot x_{\mu} +q^{1/2} i \chi \psi_{\mu})
[(\dot x +q^{1/2} i \chi \psi)^2 ]^{-1/2} \;m,
\label{24}
\end{equation}
where the following chain rule has been used
\begin{equation}
\frac{\partial L_{0}^m}{\partial
\dot x^{\mu}} = \frac{\partial (\dot x +q^{1/2} i \chi \psi)^2}
{\partial \dot x^{\mu}}\;
\frac{\partial[ (\dot x +q^{1/2} i \chi \psi)^2]^{1/2}}
{\partial (\dot x+ q^{1/2} i \chi \psi)^2}\;
\frac{\partial L_{0}^m }{\partial [(\dot x + q^{1/2} i \chi \psi)^2]^{1/2} }.
\label{25}
\end{equation}
In the computation of the $q$-derivative of the $q$-variables
with fractional power, one has to use
\begin{equation}
\frac{\partial}{\partial y} ( y^{r/s} ) = \;
\frac{( 1 - q^{2r})}{(1 - q^{2s})} \;\;y^{(r/s) - 1}, \label{26}
\end{equation}
where $r$ is {\it not} divisible by $s$ ($ r,s \in {\cal Z} $). The other
constraint: $ p\cdot \psi - m\;\psi_{5} = 0$ emerges due to
$\frac{\partial L_{0}^m}{\partial\chi} = 0$. The latter constraint and the
mass-shell condition (19) are satisfied for both the left chain rule as well
as the right chain rule of differentiation.

It is a well established fact that the existence of first-class constraints on
a system implies underlying gauge symmetries [14]. For the system under
consideration, there exist first-class as well as second-class constraints
which can be seen for all the three equivalent
Lagrangians (11), (15) and
(22). The primary constaints
momenta $ \Pi_{e} \approx 0$ as well as  $ \Pi_{\chi} \approx 0  $
and, corresponding secondary constraints
$ p^2 - m^2 \approx 0,\; p \cdot \psi- m\psi_{5} \approx 0 $, are first-class.
However,
the canonical momenta corresponding to the fields $ \psi_{\mu}$
and $\psi_{5} $
are second-class. We shall not devote time here for the discussion of
$q$-deformed Dirac brackets, subsequent $\hbar$-quantization schemes, etc.,
in the Hamiltonian formalism.
Instead, we shall concentrate {\it only} on the first-order Lagrangian (11)
and discuss local (super)gauge symmetries as well as reparametrization
invariance that emerge due to the presence of the first-class constraints. For
instance, the constraints $\Pi_{e} \approx 0 $ and $ p^2-m^2 \approx 0$
produce the following infinitesimal local gauge symmetry transformations [14]
\begin{eqnarray}
\delta_{1} x^{\mu}&=& q^{1/2}\;\xi\;p^{\mu}, \quad \delta_{1} p^{\mu} = 0,
\quad \delta_{1} e = q^2 \; \dot \xi, \nonumber\\
\delta_{1} \psi^{\mu}&=& 0\;, \qquad \delta_{1} \psi_{5} = 0,
\qquad \delta_{1} \chi = 0,
\label{27}
\end{eqnarray}
because the Lagrangian transforms as
\begin{equation}
\delta_{1}  L_{f}^m = \frac{d}{d \tau} \;
\Bigl [ \frac{\xi ( p^2 + q^2 m^2)}{( 1 + q^2 )} \Bigr ],
\label{28}
\end{equation}
where $\xi$ is an infinitesimal non-commuting gauge parameter
($ \xi\;p_{\mu} = q\;p_{\mu}\;\xi, \; etc.$)

To remove the negative norm states from the physical spectrum (that might
be generated due to the zero component of $ \psi^\mu$) one
requires a local supergauge symmetry. The constraints $\Pi_{\chi} \approx 0 $
and $p \cdot \psi - m \psi_{5} \approx 0$ serve that purpose by
generating the following
supergauge transformations with the infinitesimal $q$-(anti)commuting parameter
$\eta(\tau) (\eta^2 = 0) $
\begin{eqnarray}
\delta_{2} x^{\mu}&=& q^{1/2}\;\eta\;\psi^{\mu}, \quad \delta_{2} p^{\mu} = 0,
\qquad \delta_{2} e = q^{1/2} (1+q^2) \eta \chi, \nonumber\\
\delta_{2} \psi^{\mu}&=& q^{1/2}i\eta\;p^{\mu}, \quad
\delta_{2} \psi_{5} = q^{1/2}i\eta\;m,
\qquad \delta_{2} \chi = i\; \dot \eta,
\label{29}
\end{eqnarray}
where $\eta$ obeys following $q$-(anti)commutation relations
\begin{eqnarray}
\eta\;x_{\mu}&=&q^{1/2} \;x_{\mu}\;\eta, \qquad
\eta\;p_{\mu}=q^{1/2} \;p_{\mu}\;\eta,\qquad
\eta\;\psi_{\mu}=-q^{1/2} \;\psi_{\mu}\;\eta, \nonumber\\
\eta\;\psi_{5} &=& -q^{1/2} \;\psi_{5}\;\eta ,\qquad
\eta\;m\;= q^{1/2} \;m\;\eta ,\qquad
\eta\;\chi=-\;\chi\;\eta .\label{30}
\end{eqnarray}
As a consequence of the above symmetry transformations, the first-order
Lagrangian transforms as
\begin{equation}
\delta_{2}  L_{f}^m = q^{1/2}\;\frac{d}{d \tau} \;
\Bigl [ \frac{\eta ( p\cdot \psi + m \psi_{5})}{2} \Bigr ].
\label{31}
\end{equation}
The above Lagrangian is also
invariant under the following reparametrization transformations
\begin{eqnarray}
\delta_{r} \; x_{\mu}&=& \;\epsilon \;\dot x_{\mu}, \qquad
\delta_{r} \; p_{\mu} = \;\epsilon \;\dot p_{\mu},  \qquad
\delta_{r} \; e = \; \frac{d}{d \tau} ( \epsilon e ),\nonumber\\
\delta_{r} \; \psi_{\mu}&=& \;\epsilon \;\dot \psi_{\mu}, \qquad
\delta_{r} \; \psi_{5} = \;\epsilon \;\dot \psi_{5},  \qquad
\delta_{r} \; \chi = \; \frac{d}{d \tau} ( \epsilon \chi ),
\label{32}
\end{eqnarray}
emerging due to the one-dimensional diffeomorphism $\tau \rightarrow \tau
- \epsilon (\tau)$ (with the commuting infinitesimal
parameter $\epsilon (\tau)$), because the
Lagrangian undergoes the following change under (32):
\begin{equation}
\delta_{r} L_{f}^m = \frac{d}{d\tau} \Bigl [ \epsilon \; L_{f}^m \Bigr ].
\label{33}
\end{equation}
In the usual undeformed
($q = 1 $) case of a free spinning relativistic particle, a linear
combination of (super)gauge
symmetries (29) and (27) is found to be equivalent on-shell to the
reparametrization
invariance (32) with the
identification $\xi = \epsilon e $ and
$\eta = \epsilon \chi $ [15]. However, in the deformed case, as it turns out,
even for the above identification of the (super)gauge parameters and
the on-shell condition (13), the following equality
\begin{equation}
(\delta_{1} - i \delta_{2}) \Phi = \delta_{r} \Phi,
\label{34}
\end{equation}
for the variables $ \Phi \equiv x_{\mu}, p_{\mu}, \psi_{\mu},
\psi_{5}, e ,\chi $
is true {\it only} for $q=\pm1$. This is  due to the fact that
the transformations of the einbein field, in spite of the
above identification, are not equal on-shell unless $q=\pm1$.
This condition ($ q = \pm 1$) also turns up due to the requirement that the
$q$-commutator of two supersymmetric gauge transformations must produce a
reparametrization plus an additional supersymmetric gauge transformation
as is essential in pure supergravity theories.
It is not difficult to check that following equalities
\begin{eqnarray}
&&[ \delta_{\kappa}, \delta_{\eta} ]_{q^2}\;x^\mu = \epsilon \dot x^\mu
+ q^{1/2} \eta^{\prime} \psi^\mu, \nonumber\\
&&[ \delta_{\kappa}, \delta_{\eta} ]_{q^2}\;\psi^\mu = \epsilon \dot \psi^\mu
+ q^{1/2} i \eta^{\prime} p^\mu, \nonumber\\
&&[ \delta_{\kappa}, \delta_{\eta} ]_{q^2}\;\;\psi_{5} = \epsilon \dot \psi_{5}
+ q^{1/2} i \eta^{\prime} \;m, \nonumber\\
&&[ \delta_{\kappa}, \delta_{\eta} ]_{q^2}\;\;\chi \;= \epsilon \dot \chi
+\dot \epsilon \chi+ i\;\dot \eta^{\prime},  \nonumber\\
&&[ \delta_{\kappa}, \delta_{\eta} ]_{q^2}\;\;p^\mu = \epsilon \dot p^\mu ,
\label{35}
\end{eqnarray}
are true {\it on-shell} for
\begin{eqnarray}
&&\epsilon = i q^{1/2} (1+q^2) \eta  \kappa e^{-1}, \nonumber\\
&&\eta^{\prime} = i\epsilon \chi \equiv - q^{1/2} (1+q^2) \eta
\kappa e^{-1} \chi,
\nonumber\\
&&[ \delta_{\kappa}, \delta_{\eta} ]_{q^2}
\equiv \delta_{\kappa}\;\delta_{\eta} - q^2\; \delta_{\eta}\;\delta_{\kappa}.
\label{36}
\end{eqnarray}
However, the validity of a similar equality in the case of the einbein field,
namely, $ ( \delta_{\kappa}\;\delta_{\eta} - q^2\;
\delta_{\eta}\;\delta_{\kappa})e =\dot \epsilon \;e+ \epsilon \dot e+ q^{1/2}
(1+q^2) \eta^{\prime} \chi$, requires
\begin{equation}
( \eta \dot \kappa - q^2 \kappa \dot \eta )
= \eta \kappa e^{-1} \dot e ( 1- q^{-2} ) + ( \eta \dot
\kappa - \kappa \dot \eta ),
\label{37}
\end{equation}
which is true only for
$ q = \pm 1$.  \footnote{ In the computation of $\dot \epsilon $, we have used
$\dot e^{-1} = (\partial e^{-1} / \partial e) \dot e \equiv -
q^{-2}\;e^{-2}\dot e $. Here $\eta$ and $\kappa$ are supersymmetric
transformation parameters and the transformation $ \delta_{\kappa} [ \eta
\psi_{\mu} = - q^{1/2} \psi_{\mu} \eta ] $ leads to $ \eta \kappa = - \kappa
\eta $ due to $ \eta p_{\mu} = q^{1/2} p_{\mu} \eta $. } This only
demonstrates that for arbitrary value of $q$, the supergravity requirement and
the on-shell equivalence of (super)gauge and reparametrization symmetries are
not true.

One can compute the conserved charges corresponding to the symmetry
transformations (27) and (29) by applying the least action
principle. This is illustrated below:
\begin{equation}
\delta S = 0 = \int d\tau \Bigl (\delta
[ q^{1/2} p_{\mu} \dot x^\mu + \frac{i}{2} \psi \cdot \dot \psi
- \frac{i}{2} \psi_{5}  \dot \psi_{5} - {\cal H}_{c}( x,p,\psi,\psi_{5}) ]
- \frac{d}{d\tau} g(\tau)\Bigr ),
\label{38}
\end{equation}
where $S$ is the action corresponding to the Lagrangian (15),
$ {\cal H}_{c} $ is
the canonical Hamiltonian and
$ g(\tau) = \frac{\xi (p^2 + q^2 m^2)}{1+ q^2} \;\mbox{and}\;
\frac{q^{1/2}\eta (p\cdot \psi +  m\psi_{5})}{2} $, respectively, for the
above symmetry transformations. Using
anticommutation relations for $\psi_{\mu}$ , $\psi_{5}$ and
$q$-commutation relations $ \delta \dot x^\mu p_{\mu}
= q p_{\mu} \delta \dot x^\mu $,
we obtain Hamilton equations of motion and conservation laws.
For the validity of the following Hamilton equations of motion
\begin{equation}
\dot x^\mu = q^{-1/2} \frac{\partial {\cal H}_{c}}{\partial p^\mu}, \;
\dot p^\mu = -q^{1/2} \frac{\partial{\cal H}_{c}}{\partial x^\mu}, \;
\dot \psi^\mu = -i \frac{\partial {\cal H}_{c}}{\partial \psi^\mu},  \;
\dot \psi_{5} = i \frac{\partial {\cal H}_{c}}{\partial \psi_{5}} ,
\label{39}
\end{equation}
we obtain a general expression for the conserved charge ($Q$) as:
\begin{equation}
Q =  q^{-1/2} \delta x^\mu p_{\mu} + \frac{i}{2} \delta \psi_{5} \psi_{5}
- \frac{i}{2} \delta \psi_{\mu} \psi^{\mu} - g(\tau).
\label{40}
\end{equation}
In the case of the global version of
symmetry transformations (27) and (29), this yields the following charges:
\begin{equation}
Q_{\xi} = \frac{q^2 (p^2 - m^2)}{1 + q^2} \quad \mbox{and} \quad
Q_{\eta} = q^{1/2} ( p\cdot \psi - m\psi_{5} ).
\label{41}
\end{equation}
One can readily check that both of these
charges are conserved due to the equations of motion (13).
The latter one is conserved on the constrained
submanifold where the first-class constraint $ p^2 - m^2 = 0 $ is valid.

It is rather difficult to
extract out a general solution for equation (13) when all the variables
are arbitrary functions of the evolution parameter $ \tau $. However,
due to gauge symmetry transformations (27) and (29), one  can
choose an
analogue of the Lorentz gauge,
namely; $ \dot e = \dot \chi = 0$.
Under such a gauge choice, one obtains
\begin{eqnarray}
x_{\mu}(\tau)&=& x_{\mu}(0) + q^{1/2} [ e(0) p_{\mu}(0)
- i \chi (0) \psi_{\mu} (0)] \;\tau ,\nonumber\\
\psi_{\mu}(\tau)&=& \psi_{\mu} (0)
+q^{1/2} \;\chi (0) p_{\mu}(0)\;\tau, \nonumber\\
\psi_{5}(\tau)&=& \psi_{5}(0)  +q^{1/2} \;\chi(0)\;m \;\tau,\nonumber\\
p_{\mu}(\tau)&=&\;p_{\mu}(0),
\label{42}
\end{eqnarray}
which satisfy all the $q$-(anti)commutation relations (2), the
$GL_{q}(2)$ invariant quantum world-line
$ x_{\mu}(\tau) p^{\mu}(\tau) = q \;p_{\mu}(\tau) x^{\mu}(\tau) $
and the $q$-super world-line (3) at any arbitrary value of the evolution
parameter $\tau$, if we assume the validity of relations (2) and (12) at the
initial "time" $\tau = 0$.

Unlike the $q$-dependent (anti)commutation relations among the variables
in equations (2), (12) and (30),
there are
some {\it $q$-independent} (anti)commutation relations that emerge
automatically
due to (graded)associativity conditions {\it or} infinitesimal gauge
transformations on $q$-dependent relations. For instance, one can easily
see the commutativity of $e$ and $\chi$ that is present in equation (2).
This emerges due to the
on-shell condition $ \dot\psi_{\mu} = q^{1/2}\chi p_{\mu}  $
in $ e \dot \psi_{\mu}= q^{1/2} \dot \psi_{\mu} e $ with
$ e  p_{\mu}= q p_{\mu} e $. The commutativity of mass parameter $m$ and
momenta $p_{\mu}$ in equation (12) is mainly due to the mass-shell condition
which can be also checked by extracting out the expression for $p_{\mu}$ from
equations of motion (13) and using equations (2) as well as (12). More comments
about
this commutativity can be found
in Ref.[8]. The $q$-independent anticommutativity of $\eta$ and $\chi$
in equation (30) emerges due to
$\chi \delta_{2}\psi_{\mu} = - q^{1/2} \delta_{2}\psi_{\mu} \chi$ when
we use $ \chi p_{\mu} = q^{1/2} p_{\mu}\chi$.

It is now a very
interesting venture to develop a $q$-deformed BRST quantization scheme
on a quantum (super)world-line for a spinning and a scalar
relativistic particle, as they present a prototype example of an
Abelian gauge
theory. These examples would provide the simplest laboratory for the
development of $q$-deformed gauge theory, $q$-deformed Hamiltonian
formulation, $q$-deformed constraint analysis and $q$-deformed
Dirac brackets, etc., in the {\it undeformed}
Minkowski space-time manifold. It would be worthwhile to  extend these models
to the case when Minkowski space-time manifold and cotangent manifold
{\it both} are $q$-deformed. In addition, one can generalize the second-order
Lagrangian (15) to the corresponding $q$-deformed Neveu-Schwarz-Ramond model
for $q$-deformed string theory.  These are some of the issues for future
investigations.

{\it It is a great pleasure to thank A.Filippov for taking interest in this
work and M.Pillin for his private communication on the subject.}


\begin{thebibliography}{99}
\bibitem{1}  V.G.Drinfeld, {\it Quantum Groups }, Proc.Int.Cogr.
             Math., Berkeley, { \bf 1} (1986) 798.
\bibitem{2}  M.Jimbo,  Lett.Math.Phys. {\bf 10} (1985) 63, {\bf 11}
             (1986) 247.
\bibitem{3}  See, e.g., for review, S.Majid, Int.Journ.Mod.Phys.{\bf A5}
             (1990) 1.
\bibitem{4}  I.V.Volovich, Class.Quant.Grav.{\bf 4} (1987) L83,
             CERN preprint, CERN-TH-4781 (1987).
\bibitem{5}  H.S.Snyder, Phys.Rev.{\bf 71} (1947) 38,
             C.N.Yang, Phys.Rev.{\bf 72} (1947) 874,
             H.Yukawa, Phys.Rev.{\bf 91} (1953) 415.
\bibitem{6}  I.Ya.Aref'eva and I.V.Volovich,
             Phys.Lett.{\bf 268B} (1991) 179,\\ R.M.Mir-Kasimov, J.Phys.{\bf
	     A}:Math.Gen.{\bf 24} (1991) 4283,\\ J.Schwenk and J.Wess,
	     Phys.Lett.{\bf 291B} (1992) 273,\\ V.Spiridonov, Phys.Rev.Lett.
	     {\bf 69} (1992) 398,\\ S.V.Shabanov, Phys.Lett.{\bf 293B} (1992)
	     117,\\ A.P.Isaev and R.P.Malik, Phys.Lett.{\bf 280B} (1992) 219.
\bibitem{7}  I.Ya.Aref'eva and I.V.Volovich, Mod.Phys.Lett.{\bf A6}
             (1991) 893, Phys.Lett.{\bf 264B} (1991) 62,
             A.P.Isaev and Z.Popowicz,  Phys.Lett.{\bf 307B} (1993) 353.
\bibitem{8}  R.P.Malik, Phys.Lett.{\bf 316B} (1993) 257.
\bibitem{9}  See, e.g., K.Sundermeyer, {\it Constrained Dynamics}, (Lecture
notes in Physics),
	     Springer-Verlag (Berlin, 1982), R.Casalbuoni, Phys.Lett.{\bf 62B}
	     (1976) 49.
\bibitem{10} R.P.Malik,  (in preparation).
\bibitem{11} L.Brink, S.Deser, B.Zumino, P.Di Vecchia and P.Howe,
             Phys.Lett.{\bf 64B} (1976) 435, L.Brink, P.Di Vecchia and
             P.Howe, Nucl.Phys.{\bf B118} (1977) 76.
\bibitem{12} J.Wess and B.Zumino, Nucl.Phys.(Proc.Suppl){\bf B18} (1990) 302.
\bibitem{13} M.Pillin, W.B.Schmidke and J.Wess, Nucl.Phys.
             {\bf B403} (1993) 223,\\
             A.Schirrmacher, Max Planck Institute Preprint: MPI-PTh /92-92,
             {\it Quantum group, Quantum Space-time and Dirac Equation}. \\
             M.Pillin, J. Math. Phys.{\bf 35} (1994) 2804.
\bibitem{14} P.A.M.Dirac, {\it Lecture on Quantum Mechanics},
             Yeshiva University, N.Y. (1964),
             N.Mukunda, Physica Scripta {\bf 21} (1980) 783.
\bibitem{15} D.Nemeschansky, C.Preitschopf and M.Weinstein, Ann.Phys.
	     {\bf 183} (1988) 226.
\end{thebibliography}
\end{document}